\documentclass[fleqn,12pt,twoside]{article}
\usepackage{espcrc1}
\usepackage[latin1]{inputenc}

\usepackage{graphicx}
\usepackage[figuresright]{rotating}

\newcommand{\AmS}{{\protect\the\textfont2
  A\kern-.1667em\lower.5ex\hbox{M}\kern-.125emS}}
\DeclareMathAlphabet{\bi}{OML}{cmm}{b}{it}

\title{Multifragmentation and phase transition for hot nuclei: recent progress}

\author{INDRA and ALADIN Collaborations\\
        B.~Borderie\address[IPNO]{Institut de Physique Nucl\'eaire, 
                 CNRS/IN2P3, Univ. Paris-Sud 11, Orsay, France},
	E.~Bonnet{\addressmark[IPNO]}
                  \thanks{Present address: GANIL, DSM-CEA/CNRS-IN2P3,
                  Caen, France}, 
	F.~Gulminelli\address[LPC]{LPC Caen, CNRS/IN2P3, ENSICAEN, 
                 Univ. de Caen, Caen, France},	 
        N.~Le~Neindre{\addressmark[IPNO]}
	             \thanks{Present address: LPC Caen, CNRS/IN2P3, ENSICAEN, 
                 Univ. de Caen, Caen, France},
	D.~Mercier{\addressmark[LPC]}
	         \address[IPNL]{Institut de Physique Nucl\'eaire, 
        IN2P3-CNRS et Universit\'e, Villeurbanne, France},
	S.~Piantelli{\addressmark[IPNO]}
	             \thanks{Present address: Sezione INFN, 
                 Sesto Fiorentino (Fi), Italy},	 
	Ad. R. Raduta{\addressmark[IPNO]}
	          \address[NIPNE]{ National Institute for Physics and Nuclear
	          Engineering, Bucharest-M\u{a}gurele, Romania},	 
	M. F. Rivet{\addressmark[IPNO]},
	B.~Tamain{\addressmark[LPC]},
	R.~Bougault{\addressmark[LPC]},
	A.~Chbihi\address[GANIL]{GANIL, DSM-CEA/CNRS-IN2P3, Caen
        cedex, France},
	R.~Dayras\address[IRFU]{IRFU/SPhN, CEA Saclay, Gif sur Yvette, France}, 
	J.~D.~Frankland{\addressmark[GANIL]},
        E.~Galichet{\addressmark[IPNO]},
	F. Gagnon-Moisan{\addressmark[IPNO]}
	\address[LAVAL]{Laboratoire de
        Physique Nucl\'eaire, Universit\'e Laval, Qu\'ebec, Canada},
        D.~Guinet{\addressmark[IPNL]},
        P.~Lautesse{\addressmark[IPNL]},
	J.~Lukasik\address[IFJ]{Institute of Nuclear Physics IFJ-PAN,
        Krak\'ow, Poland},
	M.~P\^arlog{\addressmark[LPC]}
	{\addressmark[NIPNE]},
	E.~Rosato\address[NAP]{Dipartimento di Scienze Fisiche e Sezione INFN,
                  Universit\'a di Napoli, Napoli,
                  Italy},
	R.~Roy{\addressmark[LAVAL]},	  
        M.~Vigilante{\addressmark[NAP]}
        and
	J.~P.~Wieleczko{\addressmark[GANIL]}}
       
\begin{document}

\maketitle

\begin{abstract}
Recent important progress on the knowledge of
multifragmentation and phase transition for hot nuclei,
thanks to the high detection quality of the INDRA array, is reported. It concerns i) the
radial collective energies involved in hot fragmenting
nuclei/sources produced in central and semi-peripheral collisions and their
influence on the observed fragment partitions, ii) a better
knowledge of freeze-out properties obtained by means of a simulation based
on all the available experimental information and iii) the quantitative study
of the bimodal behaviour of the heaviest fragment distribution for
fragmenting hot heavy quasi-projectiles which allows the extraction, for the
first time, of an estimate of the latent heat of the phase transition.
\end{abstract}

\section{Introduction}
Nucleus-nucleus collisions at intermediate energies offer various possibilities to
produce hot nuclei which undergo a break-up into smaller pieces, which is
called multifragmentation. The measured fragment properties are 
expected to reveal and bring information on a phase transition for hot
nuclei which was earlier theoretically predicted for nuclear
matter~\cite{I46-Bor02,WCI06,Bor08}. 
By comparing in detail the properties of fragments emitted
by hot nuclei formed in central (quasi-fused systems (QF) from
$^{129}Xe$+$^{nat}Sn$, 25-50 AMeV)
and semi-peripheral collisions (quasi-projectiles (QP) from 
$^{197}Au$+$^{197}Au$, 80 and 100 AMeV),
i.e. with different dynamical conditions for their formation,
the role of radial collective
energy on partitions is emphasized and the relative importance of the 
different collective energies is extracted~\cite{I69-Bon08} (section 2).
Then, in section 3,
freeze-out properties of multifragmentation events produced in central
collisions ($^{129}Xe$+$^{nat}Sn$) are estimated~\cite{I66-Pia08} and confirm the existence of
a limiting excitation energy for fragments around 3.0-3.5 MeV per nucleon.
The deduced freeze-out volumes are used as a calibration to 
calculate freeze-out volumes for QP sources; thus one can locate where the different sources
break in the phase diagram.
Finally, in section 4, the charge distribution of the heaviest fragment
detected in the decay of QP sources is observed to be bimodal.
This feature is expected as a generic signal of phase transition in
nonextensive systems as finite systems. For the first time an estimate
of the latent heat of the transition is also extracted~\cite{I72-Bon09}.
\section{Fragment partitions and radial collective energy}
To make a meaningful comparison of fragment properties which can be 
related to the phase diagram, hot nuclei showing, to a
certain extent, statistical emission features must be selected.
For central collisions (QF events) one selects complete and
compact events in velocity
space (constraint of flow angle $\geq 60^{\circ}$).
For peripheral collisions (QP subevents) the selection method
applied to quasi-projectiles minimizes the contribution
of dynamical emissions by imposing a compacity of fragments in velocity space.
Excitation energies of the different hot nuclei produced are calculated
using the calorimetry procedure (see~\cite{I69-Bon08} for details). 
By comparing the properties of selected sources on the same excitation
energy domain significant differences are observed above 5 AMeV on both mean fragment
multiplicities, $<M_{frag}>$, even normalized to the sizes of the sources
which differ by about
20\% for QF and QP sources, and generalized asymmetry:
$ A_{Z}= \sigma_{Z} / (\langle Z \rangle \sqrt{M_{frag}-1})$.
QF sources have larger normalized mean fragment multiplicities and lower
values for generalized asymmetry. 
An explanation of those experimental results concerning fragment partitions
is possibly related
to the different dynamical constraints applied to the hot nuclei produced:
a compression-expansion cycle for central collisions and a more gentle friction-abrasion
process for peripheral ones. 

\begin{figure}[htb]
\begin{minipage}[c]{.45\textwidth}
\centering
\includegraphics[width=1.15\textwidth]
{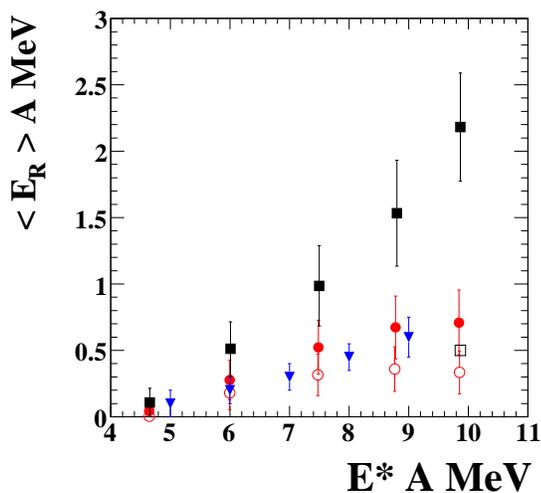}
\end{minipage}
\hspace{.05\textwidth}%
\begin{minipage}[c]{.45\textwidth}
\centering
\caption{Evolution of the radial collective
energy with the excitation energy per nucleon for different sources. Full squares stand
for QF sources. Open (full) circles correspond to QP sources
produced in 80 (100) AMeV collisions. Full triangles
correspond to $\pi^{-}+Au$ reactions~\cite{Beau01} and the open square 
to an estimate of
the thermal part of the radial collective energy for Xe+Sn sources produced
at 50 AMeV incident energy.        
From~\cite{I69-Bon08}}%
\label{fig1}
\end{minipage}
\end{figure}

Radial collective energy following a compression phase is 
predicted to be present in semi-classical simulations of central collisions 
in the Fermi energy domain~\cite{Mol88,Surau89}. In experiments
it was obtained, in most of the cases, from
comparisons of kinetic properties of fragments with models.
The mean relative velocity between fragments, $\beta_{rel}$, 
independent of the reference 
frame, allows to compare radial collective energy
for both types of sources (QF or QP). 
The effect of the source size (Coulomb contribution on fragment velocities)
can be removed  by using a simple normalization 
which takes into account, event by event,
the Coulomb influence, in velocity space, of the mean fragment charge, 
$\langle Z \rangle$, on the complement of the source charge ($Z_{s}-\langle Z \rangle$):
$ \beta^{(N)}_{rel} = \beta_{rel} / \sqrt{\langle Z \rangle (Z_{s}- 
\langle Z \rangle)}$. At
an excitation energy of about 5 AMeV, the $\beta^{(N)}_{rel}$ values
corresponding to QF and QP sources are similar.
Above that excitation energy, the values for QF sources exhibit a strong
linear increase. For QP sources $\beta^{(N)}_{rel}$ slightly increases up to
9-10 AMeV excitation energy. 
That fast divergence between the values of
$\beta^{(N)}_{rel}$ for the two types of sources signals the well known 
onset of radial collective expansion for central collisions. 
In~\cite{I40-Tab03}, estimates of radial collective energy 
(from 0.5 to 2.2 AMeV) for QF sources produced by Xe+Sn collisions are 
reported for four incident energies: 32, 39, 45 and 50 AMeV.
Those estimates which were extracted from comparisons with the statistical model
SMM assuming a self similar expansion energy can
be used to calibrate the $\beta^{(N)}_{rel}$ observable
(see~\cite{I69-Bon08} for details). 
All the quantitative information concerning 
the evolution of radial energy with excitation energy for both types 
of sources 
is presented in fig.~\ref{fig1}. We have also added the 
$E_{R}$ values published by the ISIS collaboration~\cite{Beau01}
corresponding to the $\pi^{-}+$Au reactions which
provide sources equivalent to the QP ones in terms of excitation energy
range and size. The observed evolution of $E_{R}$ for such sources is almost
the same as for QP sources. For hadron-induced reactions the thermal
pressure is the only origin of radial expansion, which indicates that it is
the same for QP sources. To be fully convincing, an estimate of the 
 part of the radial collective energy due to thermal pressure 
 calculated with the EES model~\cite{Fri90} for an excited nucleus 
 identical to QF sources produced at 50~A~MeV incident energy 
is also reported (open square) in the figure~\cite{BBro96}. 
To conclude we have shown that radial collective energy is 
essentially produced by thermal pressure
in semi-peripheral heavy-ion collisions as it is in hadron-induced reactions. 
For QF sources produced in central heavy-ion collisions the contribution 
from the compression-expansion cycle becomes more and more important 
as the incident energy increases. Those observations show that the radial
collective energy does influence the fragment partitions.
\section{Freeze-out properties}
Starting from all the available experimental information of 
selected QF
sources produced in central $^{129}$Xe+$^{nat}$Sn collisions which undergo
multifragmentation, we developed a simulation to reconstruct freeze-out
properties event by event~\cite{I66-Pia08}. The method requires data with a very high degree
of completeness, which is crucial for a good estimate of Coulomb energy.
The parameters of the simulation were fixed in a
consistent way including experimental partitions, kinetic properties and
the related calorimetry. The necessity of introducing a limiting temperature
for fragments
in the simulation was confirmed for all
incident energies. This naturally leads to a limitation of their excitation
energy around 3.0-3.5 AMeV as observed in~\cite{I39-Hud03}.
 The major properties of the freeze-out configurations
thus derived are the following: an important increase, 
from $\sim$20\% to $\sim$60\%, of
the percentage of particles present at freeze-out between 32 and 45-50 AMeV 
incident energies accompanied by a weak increase of the freeze-out volume 
which  tends to saturate at high excitation energy.  
Finally, to check the overall physical coherence of the developed approach,
a detailed comparison with a microcanonical statistical model (MMM) was
done. The degree of agreement, which was found acceptable, confirms the main
results and gives confidence in using those reconstructed freeze-out events
 for further studies as it is done in~\cite{I69-Bon08}.
Estimates of freeze-out volumes for QF sources evolve, between 32 and 50 AMeV,
from 3.9 to 5.7 $V/V_0$, where $V_0$ would correspond to 
the volume of the source at normal density; 
those volumes were used to calibrate the freeze-out volumes for QP
sources (see~\cite{I69-Bon08} for details).

The deduced volumes of QP sources are found smaller than those of QF sources
(by about 20\% on the $E^*$ range 5-10 AMeV), which supports the observations 
made previously on radial collective energies: the larger the radial
collective energy, the lower the density (the larger the F.O. volume)
where multifragmentation takes place. 
\section{Bimodality of the heaviest fragment and latent heat of
the transition}
At a first-order phase transition, the distribution of the order parameter 
in a finite system presents a characteristic bimodal behaviour in the 
canonical or grandcanonical ensemble~\cite{Gul03}.
The bimodality comes from an anomalous convexity of the underlying 
microcanonical entropy~\cite{Gros02}. It physically corresponds to the 
simultaneous presence of two different classes of physical states for the same 
value of the control parameter, and can survive at the thermodynamic 
limit in a large class of physical systems subject to long-range 
interactions~\cite{LNP02}.
\begin{figure}[!hbt]
\begin{minipage}[c]{0.45\textwidth}
\centering
\includegraphics[trim = 0 0 0 1,clip,width=1.0\textwidth]
{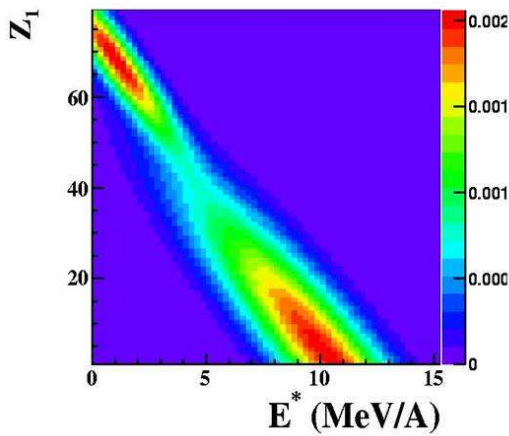}
\end{minipage}%
\hspace{.05\textwidth}%
\begin{minipage}[c]{0.45\textwidth}
\centering
\caption{
Size of the heaviest fragment versus total excitation energy in AMeV. That picture
is constructed using the fit parameters extracted from the
equivalent-canonical distribution. The distance
between the two maxima, liquid and gas peaks, projected on the excitation energy
axis corresponds to the latent heat of the transition.}
\label{fig2}
\end{minipage}
\end{figure}
In the case of hot nuclei which undergo multifragmentation, the size/charge
of the heaviest fragment was early recognized as an order
parameter~\cite{Bot01,I51-Fra05} using the universal fluctuation theory.
A quantitative analysis for QP sources is done and 
the robustness of the signal of bimodality is tested 
against two different QP selection methods~\cite{I72-Bon09}. A weighting 
procedure~\cite{Gul07} is used to test the independence of the 
decay from the dynamics of the entrance channel and to allow a comparison
with canonical expectations. Finally, a double 
saddle-point approximation is applied to extract from the measured data 
an equivalent-canonical distribution. 
To take into account the small variations
of the source size, the charge of the heaviest fragment $Z_1$ has been normalized
to the source size. After the weighting procedure, a bimodal behaviour of the largest
fragment charge distribution is observed for both selection methods.
Those weighted experimental distributions can be fitted with an
analytic function (see~\cite{I72-Bon09} for more details).
From the obtained parameter values one can estimate the
latent heat of the transition of
the hot heavy nuclei studied (Z$\sim$70) as 
$\Delta E=8.1 (\pm0.4)_{stat} (+1.2 -0.9)_{syst}$~AMeV.
Statistical error was
derived from experimental statistics and systematic errors
from the comparison between the different QP selections.
The results (for one QP source selection) are illustrated in
fig.~\ref{fig2}.


\end{document}